# Progress towards the Detailed Baseline Design for the SiD Detector Concept

Andy White for the SiD Detector Concept

University of Texas at Arlington – Physics
Arlington, TX 76019 – USA

This paper summarizes the status of the SiD Detector Concept with respect to the Detailed Baseline Design document to be prepared by the end of 2012. Each area of the SiD design is described with emphasis on the results expected for the DBD, R&D priorities, and areas of concern.

## 1 Introduction

Fig. 1 shows a general view of the SiD detector design, while Fig. 2 shows details of the subdetector locations.

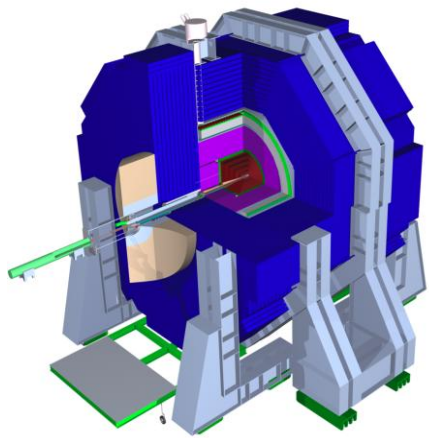
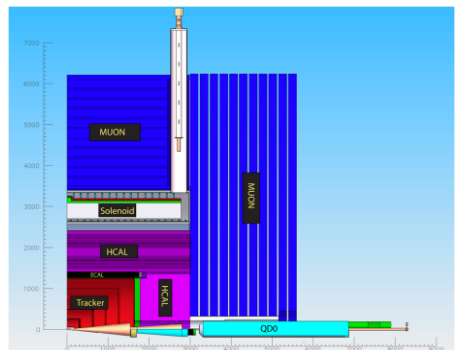

Figure 1: The SiD Detector    Figure 2: Quarter section through SiD

SiD detector is a compact, cost-constrained detector design for the International Linear Collider. It is designed to make precision measurements and be sensitive to a wide range of new phenomena. The all silicon vertexing and tracking system is the signature component of the design and immersed in a 5 Tesla field from the superconducting solenoid. Excellent momentum resolution is achieved, as is sensitivity to single bunch crossings. The calorimetry is based on the particle flow approach to achieve excellent jet energy resolution, using the high degree of longitudinal and transverse segmentation. The iron flux return, a component of the SiD self-shielding, is instrumented for muon identification and momentum measurement. The complete detector is designed for rapid push-pull operation.

The status of each subsystem of the SiD will now be discussed. More details of the components may be found in the SiD Letter of Intent [1].

## 2 Vertex detector

Fig. 3 shows the layout of the 5-layer pixel barrel and the pixel inner and outer disks.

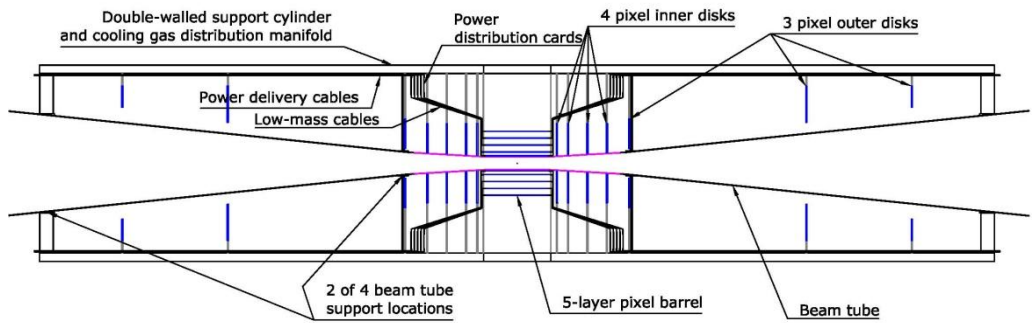

Figure 3: Vertex detector side elevation

The baseline vertex detector design uses silicon pixels arranged into an all-silicon barrel. The complete vertex system is contained within and supported by a double walled carbon fiber cylinder. There are several alternative options including silicon on foam, and silicon on carbon fiber.

There are several sensor and readout technologies being considered. Two of these, the VIP 3D chip and the Chronopixel are shown in Fig. 4.

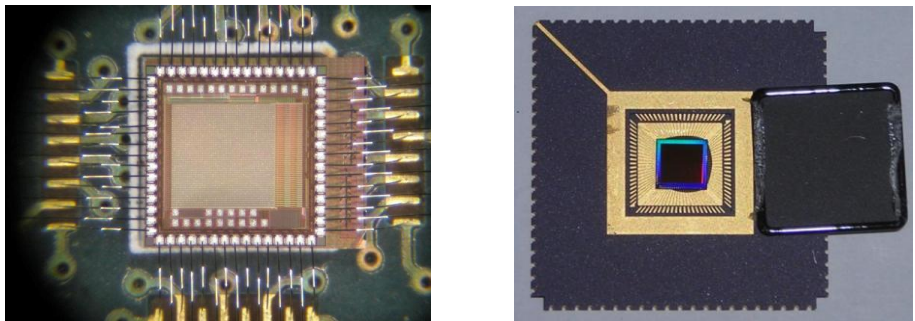

Figure 4: (left) VIP 2a chip, (right) the Chronopixel chip

VIP 2a has been produced and tested. Both analog and digital sections work. VIP 2b is in process, and sensors for 3D integration of the VIP 2b have been produced and tested. For Chronopix, the first version showed noise at design level; some issues with parts of the chip have been understood and corrected for the second prototype, which is in process and will soon be tested.

For the DBD, there will be a conceptual design for the vertex detector and integrated beam pipe design. There will be results on low-mass ladder design from the Plume collaboration. We also expect to have 3D sensor integration with the VIP 2b chip, and the results of the second version of the Chronopixel device. Funding limitations will preclude much further development of mechanical support for the vertex system.

## 3  Silicon tracking system

Figure 5 shows the arrangement of the barrel and forward silicon layers for the SiD tracking system, together with an example of the sensor tiles and readout cabling.

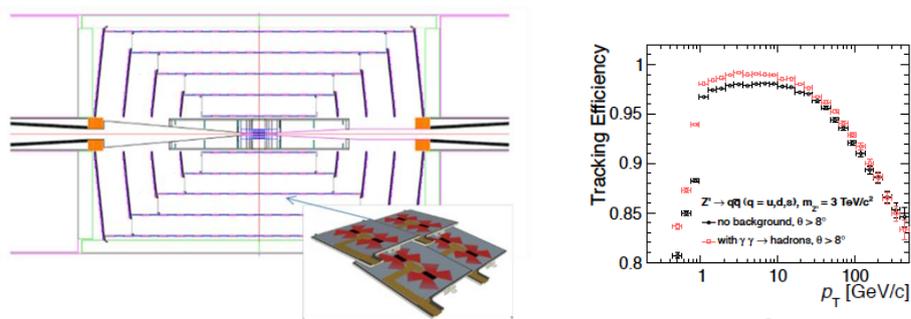

Figure 5: (left) SiD tracking system, (right) results of tracking study for CLIC_SiD

All components are in hand to produce a full silicon tracking module: sensors, the KPiX readout chip, and cabling have all been developed. Some issues have been encountered in bump bonding the chips to the sensors and commercial vendor is being consulted for possible solutions. Once the full module is assembled it will be bench tested with noise and crosstalk measurements, and, if time permits, it will be tested in the new SLAC test beam.

Tracking algorithms have been developed for the combined vertex and tracking systems and perform very well as is evidenced by the results in Figure 5 (right) for Z' to quark-antiquark with and without severe background for the CLIC_SiD detector variation.

Studies are needed for power delivery, pulsed power, and associated vibrations.

The precision alignment and monitoring is based on laser scanned interferometry which has been shown to perform at the 200nm level.

## 4  Electromagnetic Calorimeter

The electromagnetic calorimeter uses silicon sensors and tungsten absorber plates with 30 longitudinal samplings and 13 mm$^2$ transverse pixel size. This yields a small Moliere radius that promotes efficient separation of electrons, photons and charged hadron tracks. Both the electromagnetic and hadron calorimeters must allow for track imaging and reliable track/energy cluster association for use in particle flow algorithms.

Figure 6 (left) shows the layout of the electromagnetic calorimeter and its layer structure, and a view (right) of a hexagonal sensor – see Fig.6.

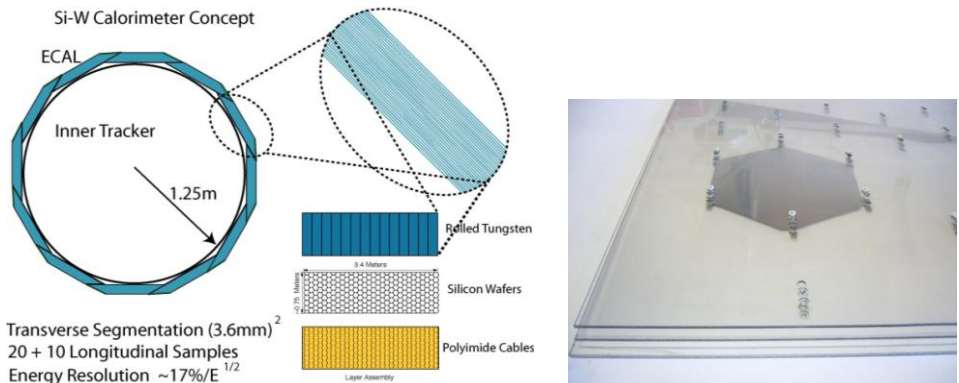

Figure 6: (left) arrangement of the electromagnetic calorimeter modules, and (right) a hexagonal silicon sensor.

The readout is based on the KPiX chip. Sufficient sensors and chips are in hand to assemble a 30-layer prototype module. Module construction awaits the resolution of the chip/sensor bonding issue described above for the tracking system. It is hoped to assemble the module and test it in the SLAC test beam in order to have results for the DBD.
An alternative technology based on monolithic active silicon pixels (MAPS) is also under consideration and would have pixel sizes of the order of 50 micron.

## 5  Hadron Calorimeter

The hadron calorimeter is mounted in azimuthal section inside the superconducting solenoid and, together with the electromagnetic calorimeter, allows the tracking of charged particles and their association with energy clusters, and the identification and measurement of the energy of energy clusters created by neutral particles. These capabilities are achieved through fine longitudinal (40 layers), and transverse (1 cm$^2$ cells) segmentation and are an essential component of the particle flow technique for jet energy measurement. The geometrical design of the hadron calorimeter is shown in Figure 7 (left). Both projective and non-projective intermodule boundaries have been considered and an initial study has shown only a small difference in energy resolution – the baseline design is therefore projective as there are benefits in having a single and simple module design.
The baseline technology for the hadron calorimeter is glass RPC's; other options include GEM's, Micromegas, and scintillator tiles. The RPC approach has been the subject of extensive testing: a 1m$^3$ module of double-glass RPC's with steel absorber plates has been built and exposed to a wide variety of particle types and momenta. Figure 7 (right) shows the 1m$^3$ RPC prototype and an example of the high degree of detail seen with this highly segmented calorimeter.

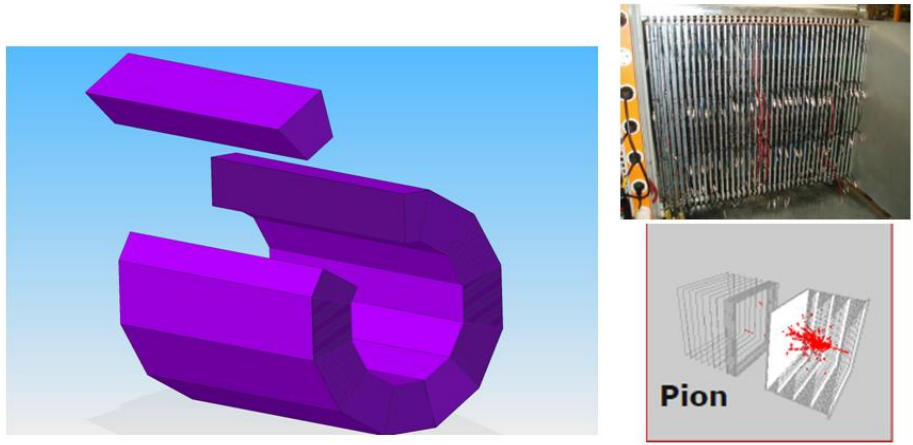

Figure 7: (left) Geometry of the hadron calorimeter, (right) RPC 1m$^3$ module and example of a pion induced shower.

For the DBD, there will be a description of the RPC-based calorimeter and its readout, results from beam tests (pions, positrons,…), calibration with muons, detailed noise measurements, and possibly results from the RPC's in a stack with tungsten plates. Further progress on the module design, high voltage and gas distribution systems will be limited by available engineering and technical resources.

## 6  Superconducting solenoid and detector integrated dipole

The 5T solenoid is based on the 4T CMS individual self-supporting winding turn design, with an integrated dipole (DID). A two-dimensional field analysis to minimize stray fields has been completed and a three-dimensional model (including DID) is in process. The analysis of the vacuum shell has been completed; work on the coil support, thermal shield and current lead design needs to be completed. The layout of the cryogenics is completed but needs integration with the QD0 quadrupole magnet. For the DBD we also expect to describe the assembly and construction procedures, and details of the power supply, dump circuit, grounding scheme, and instrumentation. Investigation of potential cost savings by the use of advanced high-purity aluminum stabilizers and conductor fabrication techniques will have to follow work for the DBD.

## 7  Muon detector system

The SiD muon system is designed to identify muons from the interaction point, and reject hadrons, with high efficiency. Figure 8 shows the layout of the steel layers for the muon system and the concept of chamber insertion for the endcap systems.

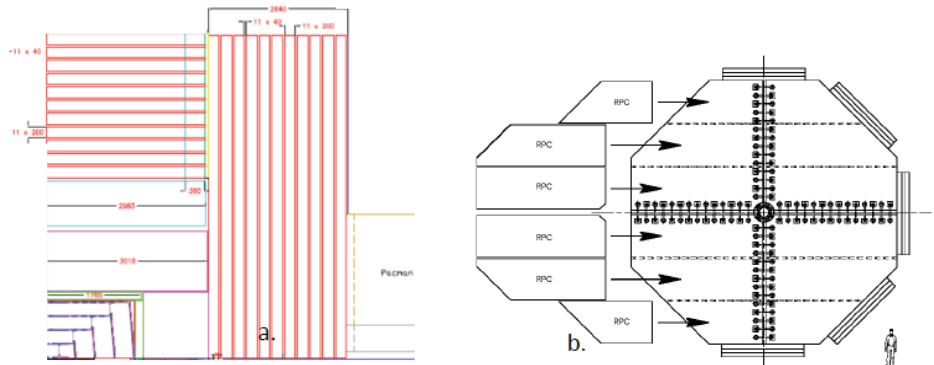

Figure 8: (left) SiD muon system layout, (right) muon endcap chamber insertion.

The baseline technology for the muon system is double layer RPC chambers, with an alternative design using long scintillator strips and wavelength shifting fibers. Aging studies with BaBar muon chambers are ongoing together with exploration of the production of new impregnated Bakelite from China, and the search for non-fluorine based gases. Results from these studies are expected for inclusion in the DBD.

# 8 Very forward region

Figure 9 shows the layout of the far forward region with the LumiCal and BeamCal systems.

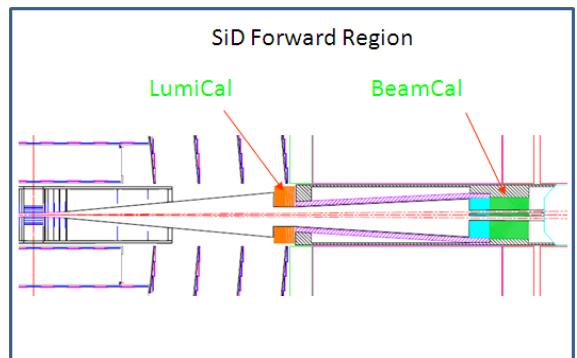

Figure 9: Schematic of the SiD forward region with Luminosity and Beam calorimeters.

The forward calorimeters provide precise measurements of the integrated luminosity and luminosity spectrum, and extend the calorimeter system down to small angles for excellent hermeticity and a two photon veto for particle searches. The calorimeters are required to be radiation hard and need fast readout due to the expected high rate. Radiation hard sensor studies are underway and initial results should be ready for the DBD. We are also following the developments of the Bean chips from the FCAL collaboration, with results from the second prototype also targeted for the DBD.

## 9 Machine-detector interface

It has been agreed that both detectors for the ILC will be mounted on platforms that will move on rollers (see Figure 10).

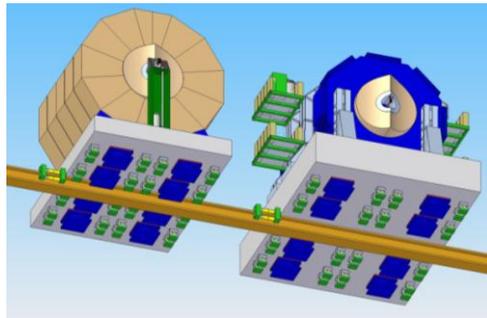

Figure 10: The SiD and ILD detectors mounted on platforms with roller motion systems.

For the DBD we expect to have a design for the push-pull motion system, design(s) for the interaction region hall, an integrated design for the beam pipe and forward calorimeters, performance results from FONT prototypes of intra-train feedback from ATF2, and, manpower permitting, results from a QD0 prototype.

## 10 Simulation and reconstruction

Since the submission of the SiD Letter of Intent, the descriptions of the subsystems have been upgraded with more realistic descriptions. Figure 11 shows an example of a simulated event in the updated geometry.

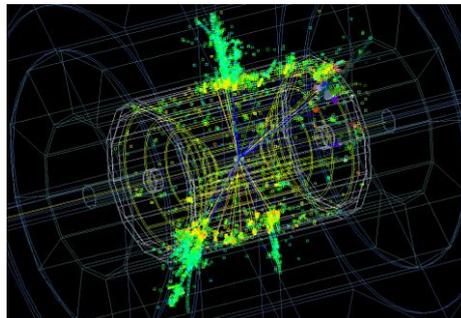

Figure 11: Simulated $e^+e^-$ collision event in the SiD detector.

The use of the SiD simulation, reconstruction, and analysis software for the CLIC CDR provided a valuable stress test, with required improvements. such as event overlay, track finding in dense environments, and the use of the PandoraPFA particle flow system, directly benefitting the DBD. We will continue to include refinements, such as more realistic calorimeter signal simulation, as they become available. We also foresee inclusion of results from the alternative IowaPFA particle flow algorithm.

## 11  Physics benchmarks

The DBD will include results from three benchmark physics processes at 1 TeV center-of-mass: $e^+e^-$ -> WW, $e^+e^-$ -> t tbar higgs, and $e^+e^-$ -> nu nubar higgs. We will also revisit one or more of the LOI benchmarks depending on available effort. Beam related background will be generated using the latest 1 TeV mechine parameters. Physics backgrounds to be simulated include 2-4-6 fermion Standard Model processes, 8-fermion backgrounds from ttbb and ttZ, $\gamma\gamma$ mini-jet events, and low-Pt, high cross-section $\gamma\gamma$ -> hadrons.

## 12  Summary

The status of each area of the SiD detector concept with respect to the DBD has been described. A team of overall and subsystem editors has been appointed for the writing of the DBD. A comprehensive outline will be produced in March 2012, the first draft of the DBD is foreseen for September 2012, with the final document available in December 2012.

## 13  References

[1] SiD Letter of Intent: https://confluence.slac.stanford.edu/display/SiD/LOI